\documentstyle[aps, epsf, floats, twocolumn]{revtex}

\begin{document}

\title{ Accurate Computation of the Magnetic Susceptibility for the Normal Phase of Organic Conductors}
\author{S. Moukouri}
\address{Department of physics, University of Cincinnati, 
Cincinnati, OH 45221}
\date{\today }
\maketitle

\begin{abstract}
The magnetic susceptibility of the quarter-filled one-dimensional extended 
Hubbard model is calculated using  the density-matrix renormalization group
technique.
  It is found that in the charge gap regime of the
model ($U> 4t $ and $V > 2t$), or in the metallic region with important  
superconductive fluctuations ($U<4t$ and $V>2t$), 
$\chi(T)$ displays a singularity at
$T=0$ and an inflection point at low temperatures that are similar to
what occurs in the spin-half quantum spin chain.  
These results, whose accuracy outdoes that of any other available
technique, are useful data which allow a comparison between theory and 
experiment in the normal phase of the organic conductors. 
\end{abstract}


During the last three decades, there has been an impressive
collection of experimental results on organic conductors
such as TTF-TCNQ, Bechgaard salts and their sulfur analogs.
These materials display various low-temperature states including 
superconductivity, itinerant and localized spin magnetism,
charge and spin density wave collective ordering, spin-peierls instability,
quantified nesting and quantum hall effects\cite{bourbonnais1}. It is suspected that
this rich phenomenology is the consequence of the combined influence
of low dimensionality and electron correlations.  These materials
are built from large planar molecules with $\pi$-orbitals that are
oriented out of the molecular plane. Intersite hopping is favored
along the stacking direction ($t_{a}$ ). The hopping integrals in the
planar directions are respectively $t_{b}=t_{a}/10$ and $t_{c}=t_{a}/500$, which
lead to strongly anisotropic electronic properties. It is generally
believed that when $T>T_{X}=t_{b}/\pi$, the physics of these materials
is mainly one-dimensional. This stems from the fact that there is 
experimental evidence that above $T_{X}$, these materials cannot be
described by the Fermi liquid theory. In particular,
it has been experimentally found that their magnetic susceptibility
 $\chi(T)$ displays a temperature dependence which is not compatible with
a Fermi liquid. Instead of the Pauli behavior of a Fermi liquid, a sizeable
decrease is observed as $T$ is lowered. 
The Luttinger liquid theory (LL),
which plays in one dimension the role of the FL, has been proposed
to interpret experimental results\cite{voit}. Many of the available LL predictions
are valid only at very low temperatures, but in this region, the 3D
character reemerges. Thus, one resorts to the direct study of model
Hamiltonians.
 
The 1D Hubbard model which belongs to the LL universality
class is believed to be the appropriate model of organic compounds.
Although the ground state properties of the basic Hubbard model( $V=0$ in
 equation (3) below) were obtained long ago\cite{lieb}\cite{shiba}, it was only recently
that its thermodynamic properties were successfully calculated \cite{juttner}. There are
however experimental constraints which imply that the basic Hubbard
model is not the appropriate model for the organic materials\cite{voit}. 
For instance, the observation of $4k_{F}$ fluctuations in TTF-TCNQ imply that
the correlation function exponent  $K_{\rho}< 1/2$.
 NMR measurements on the $(TMTSF)_{2}X$ materials predict
that $K_{\rho}=0.15$. These values are confirmed by photoemission measurements
 \cite{voit}. In the basic Hubbard model, however, $K_{\rho}$ can not
be less than $1/2$\cite{schulz}.

The discrepancy between the prediction of the basic Hubbard model and 
 the experiments, has led to the suggestion that  dimerization
and next nearest neighbor repulsion $V$ may play a non-negligible 
role\cite{schulz}\cite{voit}.
 However, when these parameters are added, the model is not solvable.
  Knowledge of the strength of correlations is of central
importance. The correlation parameters can be directly extracted from
$\chi(T)$. Fermi liquid theory and band structure calculations predict values
that differ by an order of magnitude\cite{bourbonnais1}. Current efforts are oriented
towards the computation of more reliable values of $\chi(T)$. This quantity
is very useful to determine the correlation parameters.
Bourbonnais\cite{bourbonnais2} applied a perturbative
renormalization group (RG) to the g-ology Hamiltonian, which for an appropriate
choice of parameters is a low energy limit of the Hubbard model. 
He computed $\chi(T)$ and found that its T dependence was important 
down to $T=0$. His results are, however,
 reliable only in the weak coupling regime and at low $T$.
 Mila and Penc \cite{mila1} used the world-line quantum Monte Carlo
algorithm to compute $\chi(T)$ for the dimerized Hubbard model.
 Aside from the usual statistical errors that become important at
low temperatures,  another factor that reduced the quality of their
calculation was that they had to perform three different extrapolations
on  momentum, imaginary time step and chain length.
A recent analytical RG supplemented with a diagrammatic expansion and
a QMC calculation  predicted that $\chi(T)$ would
display a singularity at $T=0$ and two inflection points at low $T$ for
the basic Hubbard model\cite{bourbonnais3}.
These features were first found in the susceptibility of the
spin-half Heisenberg chain\cite{eggert}. 
 These fine structures are, however, overlooked by QMC. The reason is
 that besides the limited accuracy, the number of points computed
in the low-temperature region remains small to allow these structures to
be resolved.

In this letter, we show that a thermodynamic density-matrix renormalization
group (DMRG) algorithm presented in an earlier publication \cite{caron1} can lead to an
accurate determination of $\chi(T)$ for the extended Hubbard model.
Our results are in excellent agreement with the exact results of 
Ref\cite{juttner} for $V=0$.
We identify two sectors corresponding to the two regimes of the
model. In the first,  where the Hamiltonian is gapless in both
charge and spin channels, $\chi(T)$ increases moderately with $V$.
But in the charge gapped regime of the model we observe an important
rise of $\chi(T)$. We reach low enough temperatures to
clearly see the appearence of an inflection point as soon as we
enter the regime where the model presents a charge gap ($V>2t$).
 At low $T$,  we find that the values of $\chi(T)$ do not smoothly extrapolate
to the zero temperature susceptibility $\chi_{0}$. This lead us to
conclude that a singularity exists at $T=0$. 

We apply the thermodynamic DMRG algorithm of Ref\cite{caron1}\cite{caron2}.
This algorithm is identical to the original $T=0$ DMRG algorithm\cite{white}. 
But the computation of the low-lying states which was the goal of
the $T=0$ algorithm is now regarded as an intermediary step. In this step,
we calculate as many target states as necessary to build a
renormalized Hamiltonian which well  describes  higher
energy states. For this, we target $M$ states belonging to different 
spin sectors.
The reduced density matrix may be written as follows,

\begin{eqnarray}
\rho =\sum_{k}\omega _{k}\rho _{k},    
\end{eqnarray}

where $\rho_{k}$ and $\omega_{k}$ are respectively the reduced
density matrix of the $k^{th}$ state and its weight. $\rho_{k}$
is given by the wave function relation

\begin{eqnarray}
\rho_{k}(i_{1},i_{2};i'_{1},i'_{2}) =
\sum_{i_{3},i_{4}}\Psi_{k}(i_{1},i_{2},i_{3},i_{4})
\Psi_{k}^{*}(i'_{1},i'_{2},i_{3},i_{4}).    
\end{eqnarray}

 As usual, the four indices represent the different blocks forming the
superblock.
 The main step of the algorithm is the exact diagonalization, by
a dense matrix technique, of the reduced superblock hamiltonian made
of the two external blocks.
 The eigenvectors obtained this way are used for
the computation of thermodynamic quantities.
A systematic thermodynamic analysis is made in the following way.
We start by fixing the temperature range. $\chi(T)$ is calculated 
for initial values of block states $m$ and target states $M$. Then
its dependence is analyzed by varying $m$ and $M$. We attribute the 
same weight to each target state, $\omega=1/M$. The importance of a target
space in the reduced density matrix is thus given by the number
of target states fixed for this space. We find that at  the
largest lattice size studied here, a good convergence is achieved
by attributing 32 states to the $S_{T}^{z}=0$ sector and 4 states to
each of the other sectors. We keep a total of 300 states in the
two external blocks and the maximum truncation error is less than
$p(m,M)<10^{-4}$. A detailed analysis of the procedure has been
given elsewhere\cite{caron2}.  $\chi_{0}=\chi(T=0)$ is calculated by the use 
of the LL relation $\chi_{0}=2/(N\Delta_{\sigma}(N))$, 
where $\Delta_{\sigma}(N)$
 is the finite size gap defined as the energy difference between
the lowest state with $S_{T}=1$ and the ground state, 
$\Delta_{\sigma}(N)=E_{1}(N)-E_{0}(N)$. We find that the calculated
$\chi_0$ is in excellent agreement with Shiba's exact result \cite{shiba}.

 The thermodynamic equations of the basic Hubbard model were
formulated long ago \cite{takahashi}, but, 
it was only recently that J\"uttner et al. succeeded
in computing physical quantities at finite $T$. In their study,
 these authors took a different approach which
avoided the difficulties of the thermodynamic Bethe ansatz equations(BAE).
 These are an infinite set of coupled nonlinear integral equations that
involve an infinite number of unknown functions.
Actual calculations were possible with only a restricted basis \cite{usuki}.
 J\"uttner et al. used the mapping of the Hubbard model to the two-dimensional
classical Shastry model. They then applied the transfer matrix method
 to derive a set of BAE with a finite number of unknowns. The resulting
nonlinear integral equations are solved numerically. The DMRG algorithm
presented above provides a possibility of making an independent verification of
their results. 
This algorithm has been quite successful in the study
of quantum spin chains\cite{caron1} and of a 
magnetic impurity in spin chains\cite{zhang}. Its
first application to fermion systems served for illustrative purposes
and a detailed analysis of its convergence was made \cite{caron2}. 
In that study, it was
mathematically shown that the finite-$T$ algorithm is as rigorous as the $T=0$
algorithm. The only difference is that there are two control parameters,
$m$ and $M$, instead of $m$ only.

 For a quantitative analysis with experimental results in the 
normal state of organic conductors, the effects of dimerization
of the bond-length in the direction perpendicular to molecular planes
should also be considered. 
In this study we consider the extended Hubbard model at quarter band-filling,
which is the density of many organic compounds,
\begin{eqnarray}
H=-t\sum_{i}(c_{i\sigma }^{+}c_{i+1\sigma }+hc)+U\sum_{i}n_{i\uparrow
}n_{i\downarrow } \nonumber \\
+V\sum_{i}n_{i}n_{i+1}.                               
\end{eqnarray}
 
The phase diagram of the model(3) has been obtained by Mila and
coworkers \cite{mila2}\cite{mila3}. They applied the exact diagonalization
on small clusters as well as analytical techniques in various
limits to show that the model has two regimes. 
The first regime is a metallic Luttinger
liquid phase that is gapless in both spin and charge channels. In the LL phase,
the dominant fluctuations are spin density waves (SDW); in the strong $V$
region they observed important superconductive fluvtuations. 
 The second regime is a charge
density wave insulator characterized by an opening of the charge gap
while the spin degrees of freedom remain gapless. The boundary between
the two regimes is given by the line going from $(U,V)=(+\infty,2)$ to
 $(4,+\infty)$.

Fig. 1 shows that the excellent agreement with BAE results seen in the study of
quantum spin chains \cite{caron1}
 can also be achieved for the Hubbard model. Our results, for $U=1$ to $16$,
were obtained by studying lattices of up to $N_{max}=24$
 and then by extrapolating
to $N \rightarrow +\infty$.
 They are in excellent agreement with the transfer matrix calculation
of J\"uttner et al. The maximum relative difference between the DMRG and the
transfer matrix results is $0.05$. This occurs at low temperatures where
the quality of the DMRG calculations is affected by finite size effect.
   The lowest temperature reached in the transfer
matrix calculation is $T=0.025$. In the DMRG, the lowest $T$, 
which is roughly given by the finite size gap $\Delta_{\sigma}(N_{max})$ depends
not only on the size of the system but also on the value of $U$. We
find that paradoxically the strong coupling regime which is hard to
study by analytical or QMC methods, is the most favorable since it
has smaller $\Delta_{\sigma}$ for a fixed size. We performed an interpolation
between $T=0$ and $T=\Delta_{\sigma}(N_{max})$. We discuss below the
reason of a particular choice of the interpolating function.

\begin{figure}%
\centerline{\epsfxsize 8cm \epsffile{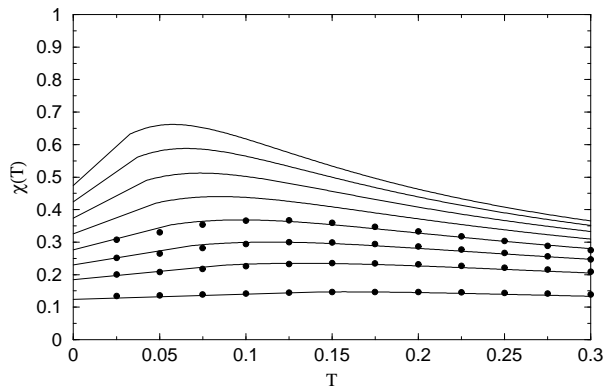}}%
\caption[]{DMRG $\chi(T)$ for $U=1,4,6,8,10,12,14,16$ from bottom to top.
The dots represent the transfer matrix results of Ref\cite{juttner}.}
\end{figure}%
                                            
A recent study \cite{bourbonnais3} has
predicted that in the weak coupling regime, (i) the position of
the maximum $T_{max}$ of $\chi(T)$ is interaction independent for
 $0 < U < 4$ , (ii) the inflection point 
at $T=0.1$ of the non-interacting model will survive and will be
interaction independent, (iii) the appearance
of a new inflection point which is the consequence of the interaction
at lower $T$, (iv) finally, the existence of a singularity at $T=0$ in
analogy with the antiferromagnetic Heisenberg spin-half chain. 
 We find that the position of $T_{max}$ is gradually moved
 toward low $T$ as the interaction increases. 
This DMRG result is more likely because in the Hubbard model, the
weak coupling regime is qualitatively similar to the strong coupling
regime. Thus one expects the same qualitative behavior in the two regions.
 We can not however directly verify the predictions (ii) and (iii) because in
the weak coupling regime, the variations of $\chi(T)$ are so small that
 these possible features may be wiped away by extrapolation
of the finite $N$ results. In the strong coupling regime, the maximum
is shifted to low T, a possible inflection point may appear at even lower $T$.
It is thus possible that we have not reached the regime predicted by the 
analytical RG.

\begin{figure}
\centerline{\epsfxsize 8cm \epsffile{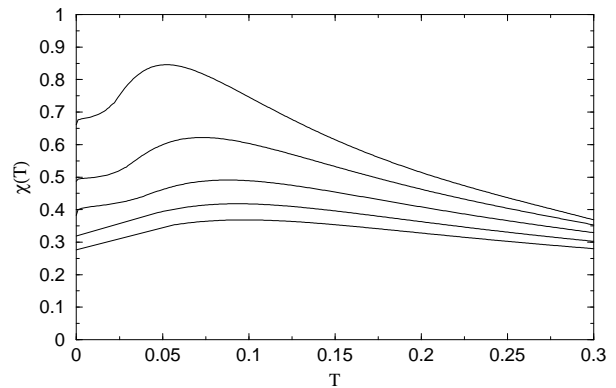}}
\caption[]{ $\chi(T)$ for $U=8$ and $V=0,1,2,3,4$ from bottom to top.}
\end{figure}   
                                           
The inclusion of $V$ in the strong coupling regime induces a clear
appearance of the regime described by the RG study. In Fig. 2 we
show $\chi(T)$ for $U=8$ and $V=0$ to $V=4$. When $U=8$ and $V=4$, we
have $\Delta_{\sigma}=0.019$. The inflection point, which is around $0.025$,
is clearly seen in our calculation. For the interpolation between 
$T=0$ and $T=\Delta_{\sigma}(N_{max})$, we tried two techniques, polynomial and
logarithmic. The latter has been tried in analogy with the study of
the spin-half Heisenberg chain \cite{eggert}. The conformal field
theory has shown that at  $T=0$, $\chi(T)$ displays a logarithmic
singularity \cite{eggert}. Since adding $V$ leads to an opening of the charge
 gap, in this regime the model bears some analogy with the spin-half
Heisenberg chain. We find that the fit $\chi(T)=\chi_{0}+A/LogT+BT$
is better than a simple polynomial fit.  The polynomial fit introduces a cusp
in $\chi(T)$. Indeed, we could also fit
the $V=0$ curves with a logarithmic function, but this seemed incorrect
 to us because we had not reached the inflection point $T_{i}$. The decrease
of $\chi(T)$ for $T >T_{i}$ is faster than for $T < T_{i}$; it is only
in this latter regime that a logarithmic fit seems correct. 
From our calculation, at first sight one should relate this low $T$
behavior of $\chi(T)$ to the onset of the insulating phase.
But the analysis below of the weak $U$ regime reveals that these
features can exist even in a metallic phase. Hence,  their occurence
in the case $V=0$ is possible.

 In the weak $U$ regime, the model remains metallic for all
values of $V$. The charge and spin degrees of freedom are both 
gapless.
We observe in Fig. 3 that for $U=2$, $\chi(T)$ increases 
 with $V$, as for $U=8$, although in this case the magnitude is smaller. For
$V=4$, we are in the region of the phase diagram where superconducting
fluctuations are important. We find that at low T, the best fit to data
 is logarithmic. Hence, the RG predictions are also true in the metallic
phase. We conclude by continuity that this is also valid when $V=0$. 
We were unable to see these features for $V=0$, either because they occur
at very low $T$ or they are wiped out by the extrapolation. The 
existence of the $T=0$ singularity may be the consequence of
 the spin-charge separation. It is therefore a generic behavior of LL
\cite{bourbonnais4}.

\begin{figure}
\centerline{\epsfxsize 8cm \epsffile{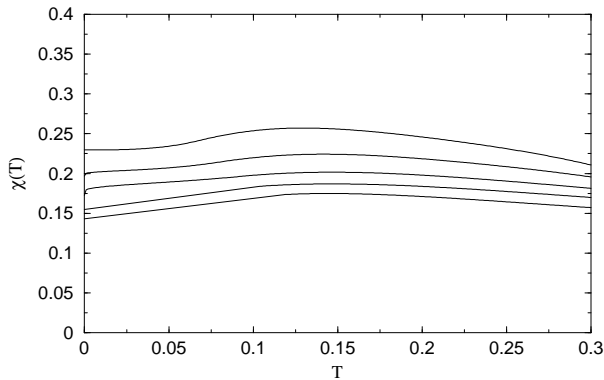}}
\caption[]{ $\chi(T)$ for $U=2$ and $V=0,1,2,3,4$ from bottom to top.}
\end{figure}   

The magnetic susceptibility is a useful quantity for the extraction
of correlation parameters $U$ and $V$ from experimental data. In
Ref\cite{bourbonnais3} it has been suggested that the ratio defined
by $R=(\chi(T_{max})-\chi(T_{i}))/\chi(T_{max})$ can be used to compare
theory and experiment. While we will perform an extensive comparison
in a longer publication, let us here look at the case of $(TMTSF)_{2}(PF)_{6}$.
In this compound, $T_{max}=275 K$ and $T_{i}=150 K$. A rough estimate
of $R_{exp}$ is $0.22$. 
It is also found
that the resistivity displays insulating behavior. This implies that
one necessarily has $U > 4 $ and $V > 2t$. In this regime, the best
match to the experimental value is obtained for $U=8$ and $V=4$. 
The corresponding estimate of the ratio is $R_{th}=0.16$. This
value can be improved either by increasing $U$ or $V$ slightly or
by including the effect of the dimerization in the Hamiltonian.
This proliferation of parameters renders any precise estimation of
$U$ and $V$ very cumbersome. It will be necessary to supplement
the susceptibility comparison with that of another physical
quantity. The rough estimation made here is in good agreement
with the values $U=7$, $V=2.8$  and $0.7$ for the dimerization
found by Mila\cite{mila4} from the analysis of the optical conductivity.
A recent reexamination of the optical conductivity by Vescoli et al.
\cite{vescoli} lead to $U=5$ and $V=2$.

In this study, we have shown  that using the finite $T$
DMRG algorithm, we can compute accurately the magnetic susceptibility
of the extended Hubbard model. We found that in the insulating regime
and in the region of dominant superconductive fluctuations, $\chi(T)$
displays a behavior that is reminiscent of the spin-half Heisenberg
chain. This is in conformity with the prediction of a recent RG study. 
 Our results can be used for extensive comparisons with experiment. 
This will be done in a
longer publication. Our current effort is the calculation of the NMR
relaxation factor $T_{1}$. A concurrent analysis of $\chi(T)$ and
$T_{1}$ will provide a better estimation of the correlation parameters.

 I whish to thank M. Jarrell for support and A. Kl\"umper for sharing
his transfer matrix data.
This work started when I was at the University of Sherbrooke (Canada). I
benefited from numerous exchanges with  C. Bourbonnais, L. G. Caron
and A.-M. S. Tremblay. 
 This work
was supported by the National Science Foundation grants DMR-0073308 and
PHY94-07194.

\end{document}